\newcommand{\msun}{\mbox{M$_{\odot}$}}
\newcommand{\ergs}{\mbox{$\rm{\,erg\,s^{-1}}$}}
\newcommand{\arcdeg}{\mbox{$^\circ$}}%
\newcommand{\arcmin}{\mbox{$^\prime$}}%
\newcommand{\arcsec}{\mbox{$^{\prime\prime}$}}%

\documentclass{iau}
\usepackage{graphicx}

\title[Light-curves of core-collapse supernovae] 
{Multi-band optical light-curve behavior of core-collapse supernovae}

\author[Brijesh Kumar]   
{Brijesh Kumar}

\affiliation{Aryabhatta Research Institute of Observational Sciences, \\ 
             Manora Peak, Nainital, 263 002, India \\ 
             email: {\tt brij@aries.res.in} }

\pubyear{2013}
\volume{296}  
\pagerange{119--126}
\jname{Supernova Environmental Impacts}
\editors{Alak Ray \& Dick McGray, eds.}
\begin{document}

\maketitle

\begin{abstract}
We present survey 
results obtained from the UBVRI optical photometric follow-up of 19 bright core-collapse SNe 
during 2002-2012 using 1-m class optical telescopes operated by the Aryabhatta Research 
Institute of Observational Science (acronym ARIES), Nainital India. This homogeneous set of data
have been used to study behavior of optical light/color curve, and to gain insight into
object-to-object peculiarity. We derive integrated luminosities for types IIP, Ibc and 
luminous SNe. Two peculiar type IIP events having photometric properties similar
to normal IIP and spectroscopic properties similar to sub-luminous IIP have been identified.  

\end{abstract}

\firstsection 
\section{Observations, data reduction and results}
One-meter class optical telescopes equipped with a modern CCD detector are best suited for optical 
follow-up of supernovae until they fade below about 20 magnitude. During 2002-2012, we monitored 
nineteen bright (peak V $<$ 16) SNe (see Table1) observable from Nainital -latitude 29\arcdeg 21\arcmin 39\arcsec N. 
The observations are carried out using 104-cm Sampurnanand Telescope (ST) at Manora Peak in 
operation since 1972 and 130-cm Devasthal Fast Optical Telescope at Devasthal in operation 
since 2010. Both these telescopes are operated by ARIES, Nainital India.
Both the telescopes are equipped with a 2kx2k CCD camera and at 104-cm ST, we have 
Johnson-Cousins UBVRI while at 130-cm, we also have SDSS ugriz filters. Both the 
telescopes are equipped with an auto-guider and the SNe observations are done in 
visitor mode. Instrumental magnitudes of SNe are estimated from profile fitting photometry 
using DAOPHOT. Landolt Standards are observed to calibrate the magnitudes of SNe. The galaxy-subtraction
technique is used for SNe embedded in the galaxy background. The light-curves (apparent \& bolometric)
are shown in Fig. 1 and the derived physical properties are given in Table 1. 

The mean of peak $M_{\rm V}$ for 6 type IIP events is $-16.7$ mag while for 6 type Ibc, 
it is $-18.0$ mag. The value of $M_{\rm V}$ for three luminous events lie between -19 
to -22 mag. Mean values of peak luminosity ($\times 10^{42}$ \ergs) are 1, 3 and 90, 
respectively for IIP, Ibc and luminous events.  
Of six IIP events, the SNe 2008in and 2012A are found photometrically similar to normal 
type IIP events, while their spectral properties suggest them to be similar to 
low-luminosity SNe. The later are thought to originate from explosion of low-mass 
8-10 \msun\ progenitors. The total radiant energy for type II SNe is of the order of $5\times10^{43}$ erg, while 
for type Ibc it is about 2 times higher. Color curve evolution of type Ibc and IIb SNe are 
quite similar.

We are thankful to the observatory staff of 104-cm and 130-cm optical telescope for their
support during observations. We gratefully acknowledge Shashi B. Pandey, Kuntal Misra, 
Rupak Roy, Brajesh Kumar, Vijay K. Bhatt for providing data and Subhash
Bose for data as well as the computer programs used in this work.

\begin{table}
  \caption{Properties of the SNe sample. The time of explosion ($t_{\rm 0}$) in
           JD 2450000+, peak V-band magnitude $V_{\rm p}$ , reddening $E(B-V)$, 
           distance modulus $\mu$, peak absolute magnitude $M_{\rm V}$, SNe types,
           peak $UBVRI/BVRI$ bolometric luminosity $L_{\rm p}$, the time when luminosity peaks,
           $t_{\rm p}$ and total radiant energy $E_{\rm t}$ integrated over the observed period 
            are given in successive columns. The reference for $t_{\rm 0}$, $V_{\rm p}$, $E(B-V)$ and SNe type is 
           given in the "Ref." column. Other parameters are derived in this work. Due to incomplete
           coverage of light-curve, the estimates for $L_{\rm p}$ and $E_{\rm t}$ are lower limits.} 
  \label{tab:sncat}
  \begin{tabular}
  {lccccclcrrr}
  \hline
  SN&  $t_{\rm 0}$ & $V_{\rm p}$& $E(B-V)$&    $\mu$&     $M_{\rm V}$&  Type&  Ref.& $L_{\rm p}$& $t_{\rm p}$& $E_{\rm t}$ \\
             &       day& mag& mag &      mag&             mag&      &      &  10$^{42}$\ergs&   day&  10$^{43}$ erg  \\ \hline  
   SN 2002ap &    2300.0&   12.42&   0.08& 29.52&  -17.35&  Ic  & 1   &     1.2&  14    & 4.4    \\
   SN 2004et &    3270.5&   12.50&   0.41& 28.70&  -17.47&  IIP & 2   &     2.2&  $<$5  & 21.0   \\
   SN 2006aj &    3784.6&   17.50&   0.15& 35.78&  -18.75&  Ibc & 3   &     5.1&  10    & 12.0   \\
   SN 2007uy &    4462.0&   15.74&   0.71& 32.37&  -18.83&  Ibc & 4   &     6.2&  22    & 16.0   \\
   SN 2008D  &    4475.1&   17.39&   0.80& 32.38&  -17.47&  Ib  & 4   &     1.6&  19    & 6.6    \\
   SN 2008gz &    4694.0&   15.50&   0.07& 32.03&  -16.75&  IIP & 5   &     0.5&  $<$87 & 2.5    \\
   SN 2008in &    4825.6&   15.07&   0.10& 30.60&  -15.84&  IIP & 6   &     0.5&  $<$5  & 2.8    \\
   SN 2009jf &    5099.5&   15.05&   0.16& 32.65&  -18.10&  Ib  & 7   &     3.0&  23    & 9.7    \\
   CSS 100217&    5170.0&   16.41&   0.02& 38.94&  -22.59&  pisn& 8   &   190.0&  $<$125& 930.0  \\
   SN 2010hq &    5395.0&   16.08&   0.08& 32.53&  -16.70&  IIP & 9   &     0.6&  $<$79 & 2.1    \\
   SN 2010jl &    5479.1&   13.80&   0.06& 33.50&  -19.89&  IIn & 9   &    19.0&  $<$28 & 200.0  \\
   SN 2010kd &    5515.0&   17.19&   0.02& 38.09&  -20.96&  pisn& 4   &    55.0&  40    & 470.0  \\
   SN 2011dh &    5712.9&   13.52&   0.04& 29.42&  -16.02&  IIb & 10  &     0.9&  $>$15 & 11.0   \\
   SN 2011fu &    5825.5&   16.92&   0.22& 34.46&  -18.22&  IIb & 11  &    38.0&  20    & 17.0   \\
   SN 2012A  &    5934.0&   13.60&   0.03& 29.96&  -16.45&  IIP & 12  &     0.6&  $<$27 &  2.8   \\
   SN 2012P  &    5930.0&   16.01&   0.05& 31.89&  -16.04&  IIb & 13  &     0.5&  $<$20 & 2.4    \\
   SN 2012aa &    5957.0&   18.00&   0.20& 37.61&  -20.78&  Ic  & 14  &    25.0&  $<$12 & 78.0   \\
   SN 2012ap &    5968.0&   16.37&   0.04& 33.45&  -17.20&  Ic  & 13  &     1.1&   11   & 2.9    \\
   SN 2012aw &    6002.6&   13.37&   0.07& 29.99&  -16.84&  IIP & 14  &     1.4&   8    & 12.0   \\
  \hline                                                                                                    
  \end{tabular}                                                                                             
  \newline\newline                                                                                             
   1-Pandey et al. (2003, MNRAS 340, 375); 2-Misra et al. (2007, MNRAS 381, 280); 3-Misra (2007, PhD Thesis);
   4-Roy \& Kumar (2013, IAUS 296); 5-Roy et al. (2011, MNRAS 414, 167); 6-Roy et al. (2011, ApJ 776, 76);
   7-Valenti et al. (2011, MNRAS 416, 318) 8-Drake et al. (2011, ApJ 735, 106); 9-Roy et al. (2011, ASInC 3, 124);
  10-Van dyk et al. (2011, ApJ 741, 28); 11-Kumar et al. (2013, MNRAS, arXiv:1301.6538 ); 12- Roy et al. (2013, IAUS 296);
  13-This work; 14-Bose et al. (2013, IAUS 296)  
\end{table}

\begin{figure}
\centering
\includegraphics[width=6.5cm]{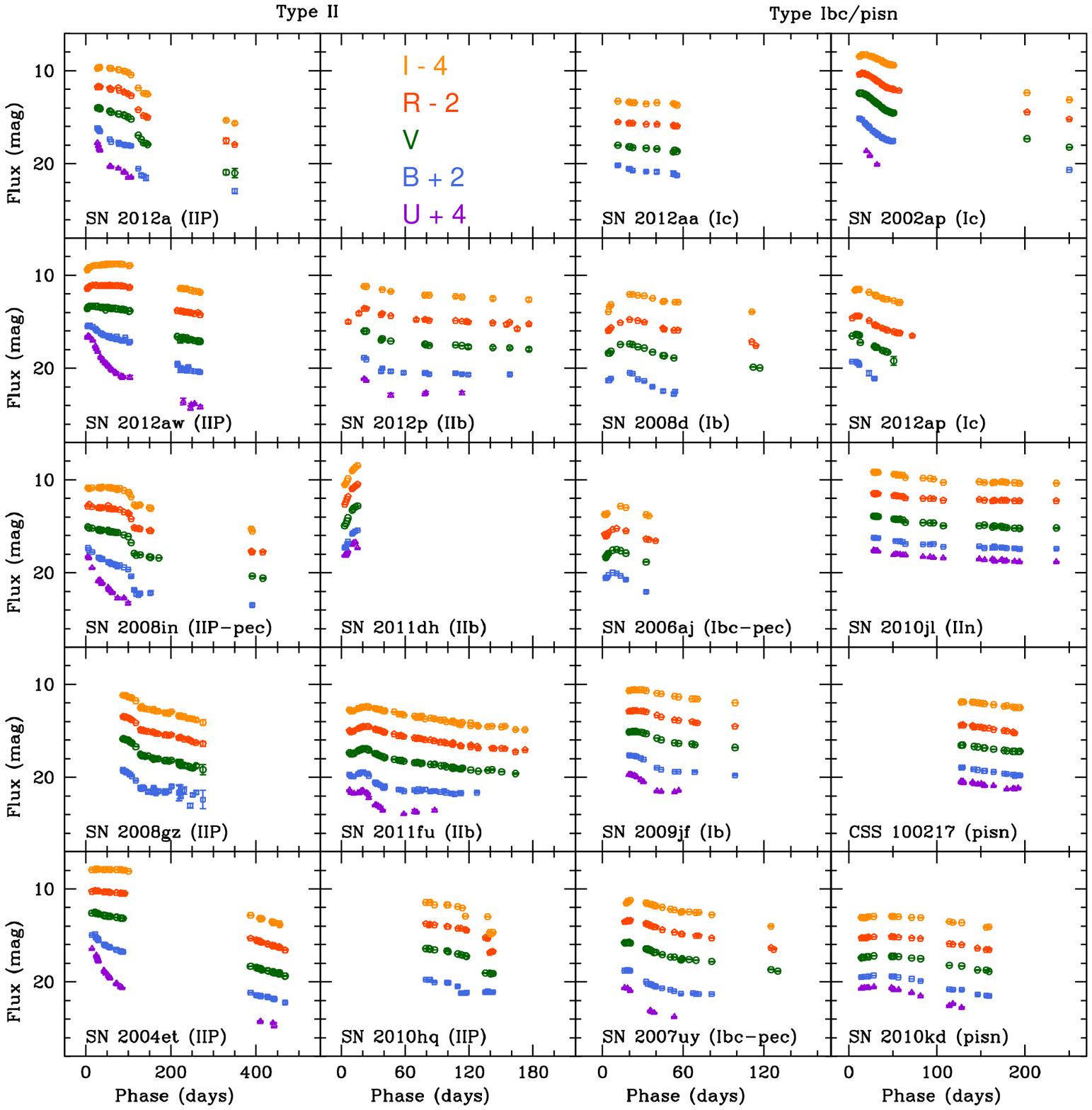}%
\includegraphics[width=6.5cm]{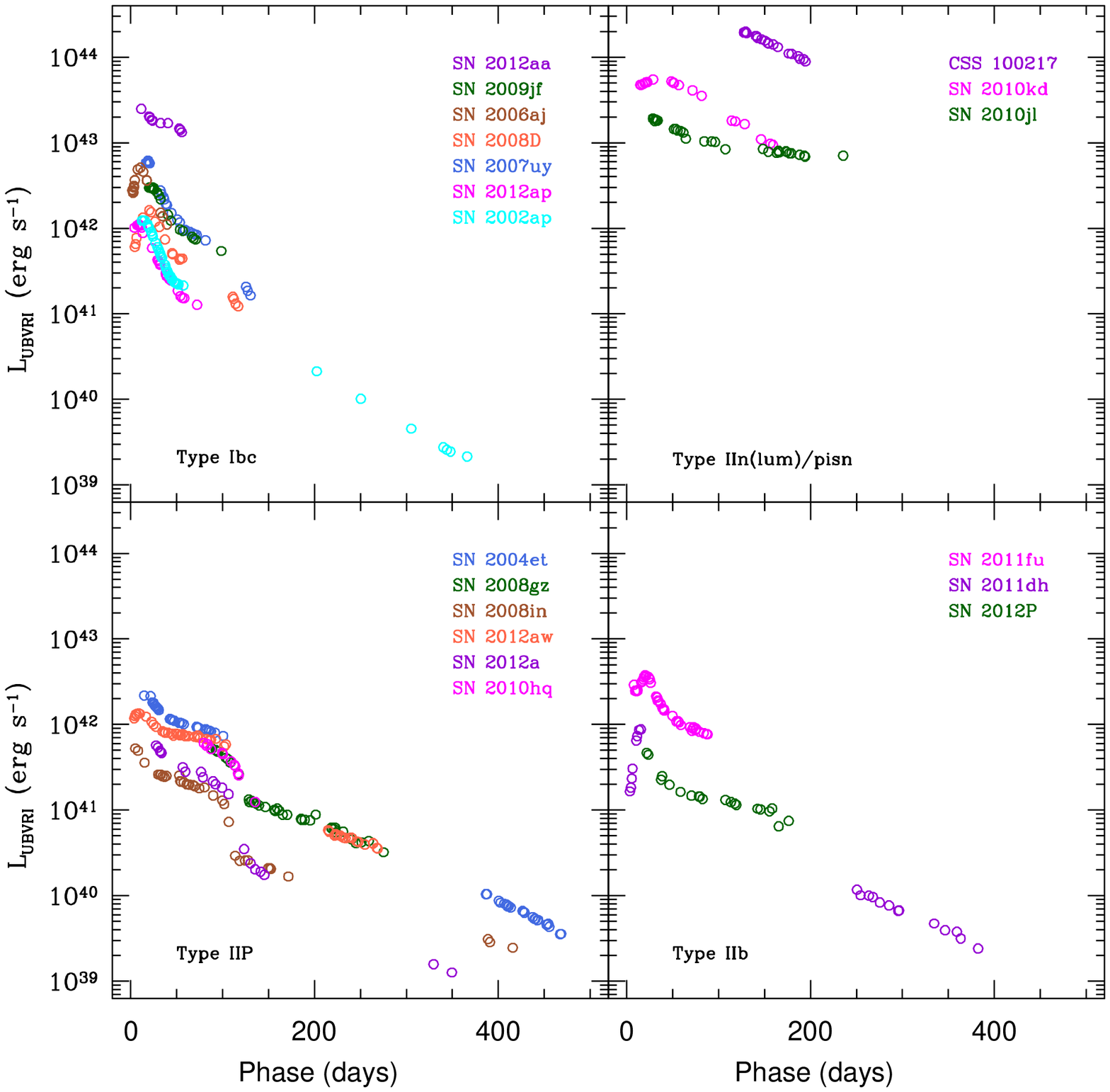}%
\caption{The $UBVRI$ apparent (left) and bolometric (right) ligh-curve of supernovae.}
\label{fig:col}
\end{figure}

\end{document}